\documentclass{article}

\usepackage[english]{babel}
\usepackage{amsmath}
\usepackage{graphicx}
\usepackage{appendix}
\usepackage{array}
\usepackage{multirow}
\usepackage{comment}
\usepackage{caption}
\usepackage{subfig}
\usepackage{booktabs}
\usepackage{authblk}
\newcolumntype{P}[1]{>{\centering\arraybackslash}p{#1}}
\usepackage{lineno,hyperref}
\modulolinenumbers[5]
\providecommand{\keywords}[1]{\textbf{\textit{Keywords---}} #1}

\UseRawInputEncoding

\begin{document}

\title{Interactive environments for training children's curiosity through the practice of metacognitive skills : a pilot study}

\author[1,2]{Rania Abdelghani\thanks{rania.abdelghani@inria.fr}}
\author[3]{Edith Law}
\author[1]{Chlo\'e Desvaux}
\author[1]{Pierre-Yves Oudeyer}
\author[1]{H\'el\`ene Sauz\`eon}
\affil[1]{Inria university of Bordeaux, Talence, France}
\affil[2]{EvidenceB, Paris, France}
\affil[3]{HCI Lab, University of Waterloo, ON, Canada}

\maketitle

\abstract{Curiosity-driven learning has shown significant positive effects on students' learning experiences and outcomes. But despite this importance, reports show that children lack this skill, especially in formal educational settings.

To address this challenge, we propose an 8-session workshop that aims to enhance children's curiosity through training a set of specific metacognitive skills we hypothesize are involved in its process. Our workshop contains animated videos presenting declarative knowledge about curiosity and the said metacognitive skills as well as practice sessions to apply these skills during a reading-comprehension task, using a web platform designed for this study (e.g. expressing uncertainty, formulating questions, etc). 

We conduct a pilot study with 15 primary school students, aged between 8 and 10. Our first results show a positive impact on children's metacognitive efficiency and their ability to express their curiosity through question-asking behaviors.}


\section{Introduction}
Curiosity plays a central role in enhancing motivation, learning and fostering academic achievement~\cite{litman2005,kang2009,jepma2012}. But despite its importance, there is evidence that curiosity decreases with formal schooling~\cite{engel2011}. 

In addressing this issue, studies such as in~\cite{jirout2018} suggest that curiosity is a malleable skill that we can explicitly promote in classrooms via training specific information-seeking behaviors. Even further, and beyond practicing such behaviors, research such as ~\cite{murayama2019,loewenstein1994,berlyne1978} suggest that training epistemic curiosity requires working on specific metacognitive skills that are involved in its process, such as the ability to evaluate and monitor one's own knowledge. 

Following these findings, the present work proposes to test a curiosity training that focuses on four specific metacognitive skills that we hypothesize are involved in its triggering and maintaining mechanisms : identifying uncertainty, generating educated guesses about it, seeking new information to resolve it and assessing the value of this new information. In practice, this training contains a first part that proposes declarative knowledge about these metacognitive skills using animated videos. The second part contains practice sessions to help participants acquire procedural knowledge about these skills. They use a dedicated web application where they interact with conversational agents that give specific cues to help them apply the said skills during a reading-comprehension task.

\section{Related work}
\subsection{Curiosity and the related metacognitive skills}
\label{section:curiosity-framework}
Looking at the relationship between curiosity-driven learning and metacognition, we find four relevant articulations : 1) the activation of epistemic curiosity mobilizes the metacognitive ability to evaluate one’s own knowledge and, more specifically, to be aware of their learning deficiencies and evaluate how close they feel to compensate for them. This is known as the information-gap theory introduced by Lowenstein in~\cite{loewenstein1994}. 2) the sustainability of epistemic curiosity states is fostered when the reward offered by the subsequent gained information is anticipated using metacognition, i.e. predictive judgment or guessing~\cite{murayama2019}.

Indeed, making educated guesses about uncertainties before looking to resolve them can amplify feelings of curiosity and guide individuals in their information-seeking behaviors in order to find specific evidence to accept or reject their hypothesis~\cite{gruber2019,gross2020}.  3) efficient curiosity-driven behaviors require metacognitive regulation skills : these latter are required for conscious information seeking, for evaluating the value of products from seeking for possible correction if unsatisfactory, and for maintaining sustained attention over long time of information search~\cite{murayama2019}. And finally 4) the individuals' capacity to monitor and assess their own learning progress is also a driver for maintaining curiosity-driven learning. This is explained in the Learning Progress (LP) Theory~\cite{oudeyer2007} suggesting that the learners become most curious when they work on tasks that provide them with an optimal level of learning progress, as perceived and evaluated by these learners themselves. Meaning that the faculty to accurately self-assess and monitor progress plays an important role in maintaining curiosity. See Appendix~\ref{appendix:cur-framework} for a view of the articulation between curiosity-driven learning and metacognition.

\subsection{Technologies for training curiosity}
Previous work has explored ways to elicit curiosity using new technology. For example, studies such as in~\cite{goren2015,ceha2019a} showed that implementing interactions with social robots that exhibit curious behaviors had significant positive effects on children's own curiosity. Similarly, previous studies suggest high benefits for using conversational agents in education to support metacognitive strategies~\cite{Grigoriadou2005} and higher-level thinking \cite{Aleven99}. Work in~\cite{alaimi2020, abdelghani2022} investigated using such agents to help children practise question asking and question-guided learning and showed rather positive results on the exploratory behaviors and subsequent domain-knowledge learning progress. 

\section{Current study}
Our present study offers a novel interdisciplinary approach which integrates psychological theories of curiosity with the design of novel educational technologies to provide a metacognitive {\it training} for curiosity. We propose an approach for testing these theories by operationalizing curiosity into a set of metacognitive exercises and studying their effects on students.

For this, we rely on the theories presented in the section above (~\ref{section:curiosity-framework}) and propose a training based on the four metacognitive skills we see involved in curiosity mechanisms : the ability to 1) identify learning uncertainties, 2) formulate educated guesses about the uncertainty, 3) lead organized information-searching behaviors to compensate for the knowledge gap and finally, 4) monitor the learning progress made through the information-seeking behavior. We call this curiosity training framework {\bf identify-guess-seek-assess}.

In practice, our training consists of two parts : the first part presents declarative knowledge about curiosity and our four metacognitive skills of interest (identify, guess, seek, assess); presented in the form of animated videos. The second part of the workshop is designed with the aim to help children acquire procedural knowledge of the 4-step curiosity-based metacognitive loop. For this, participants will practice these skills explicitly during a reading-comprehension task using a web-based platform where they interact with conversational agents that help them apply each of these skills.

We conduct a pilot study to test our training with 15 primary school students. Our evaluation is based on assessing the accessibility of our content and the pre-post intervention effects on the participants' 1) metacognitive efficiency-- which is a key facilitator for curiosity, 2) ability to ask curious questions about a task at hand and 3) their perception of the value of curiosity.

\section{Methods}
\subsection{Experimental design and procedure}
\begin{figure}[ht]
  \centering
  \includegraphics[width=1\linewidth]{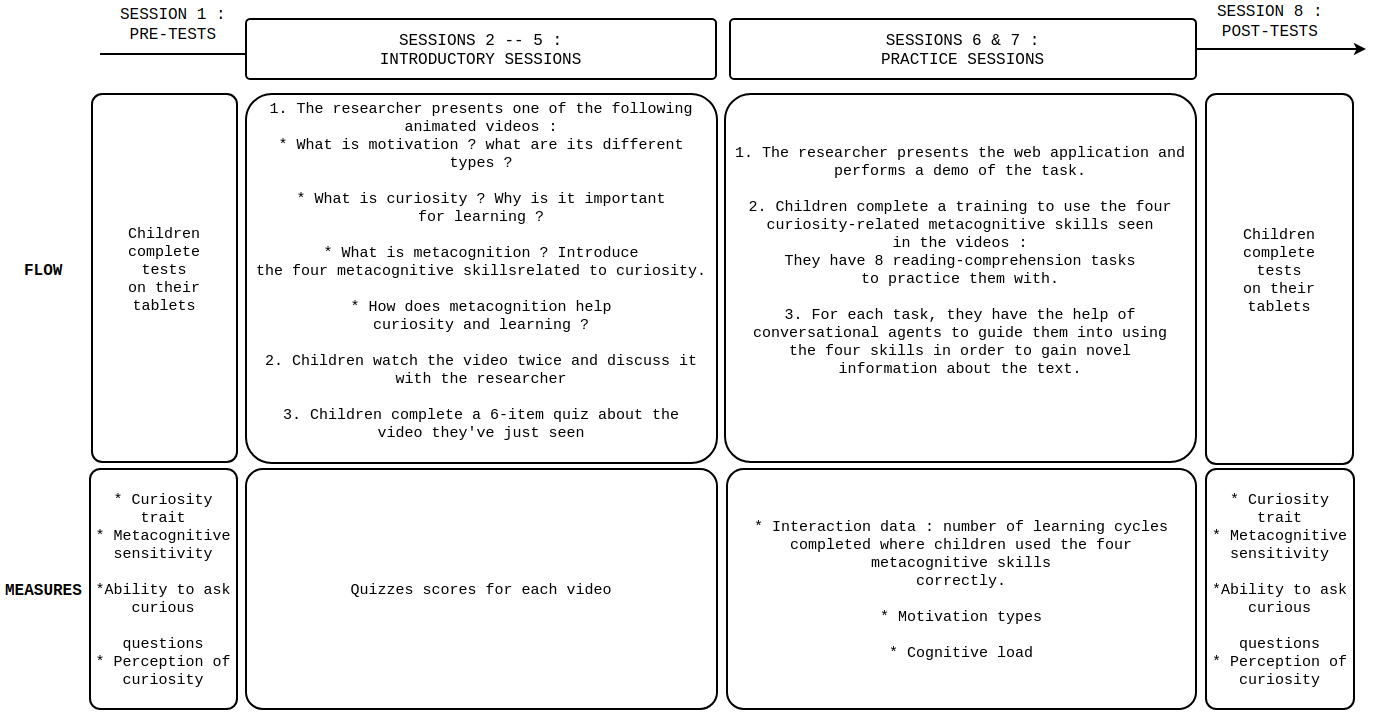}
  \caption{\textbf{Timeline of the study and associated measures}}
  \label{fig:timeline}
\end{figure}

We propose a 8-session workshop, of 45 minutes each, as detailed in Figure \ref{fig:timeline}.

\textbf{The test sessions :} sessions 1 and 8 in Figure \ref{fig:timeline} are dedicated to take the pre- and post-intervention measures related to curiosity and metacognition. Details about these measures can be found in Appendix\ref{appendix:measures}.

\textbf{The introductory sessions :} During each of these sessions (2--5), we present a 4-minute animated video explaining a concept related to curiosity and/or metacognition that we created within the team. We introduce the four curiosity-related metacognitive skills during one of these videos (see Appendix~\ref{appendix:mc-car} for more details about the content of these videos). We model the skills with 2D characters to make it easier for participants to remember their roles: a first controller to reflect the IDENTIFY skill, a detective to reflect the GUESS skill, a explorer to reflect the SEEK skill and a second controller to reflect the ASSESS skill. See Figure~\ref{fig:timeline} for more details about the procedure and Appendix~\ref{appendix:mc-car} for more details about the metacognitive characters. 

\begin{figure}[ht]
  \centering
  \includegraphics[width=1\linewidth]{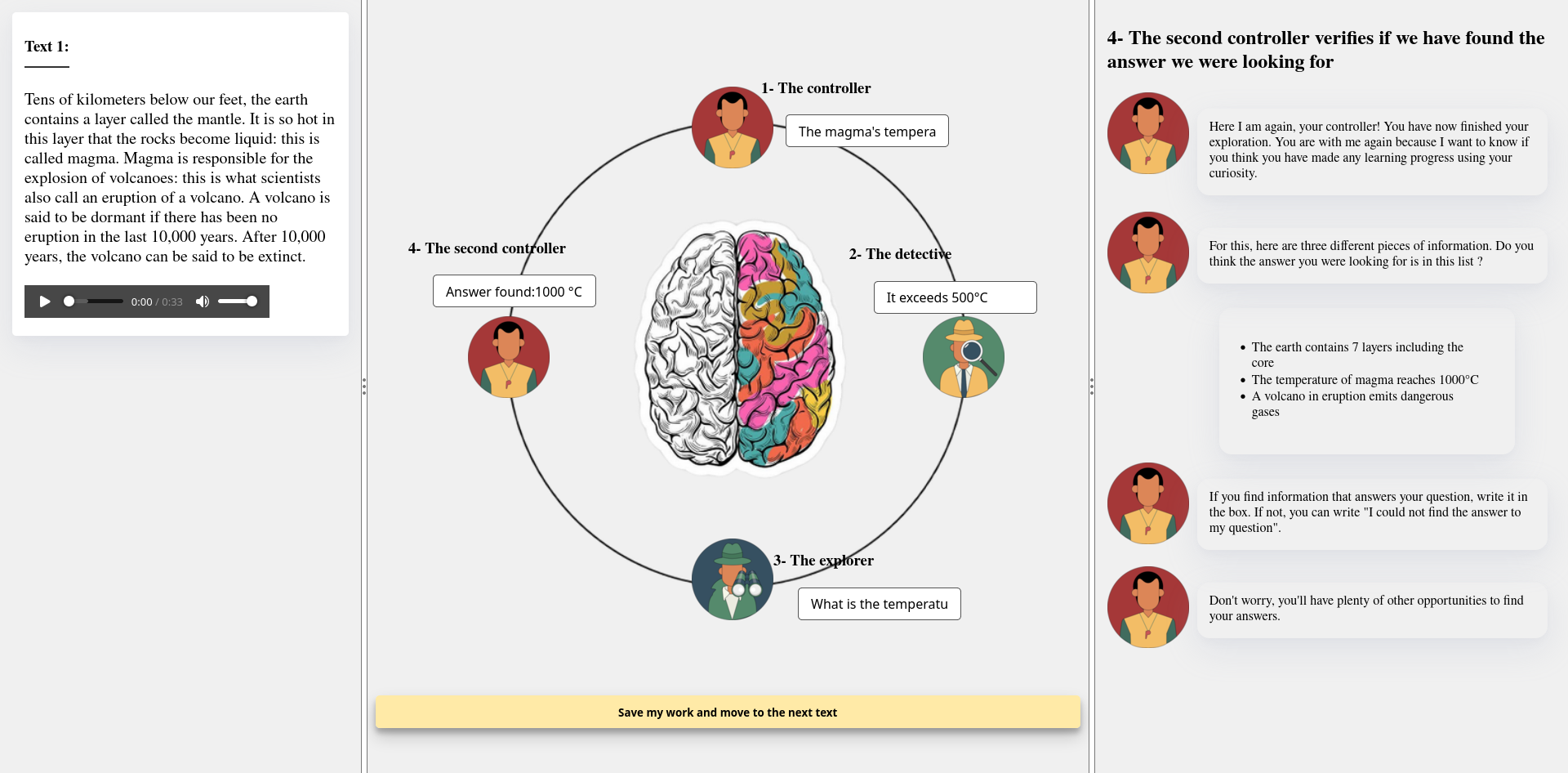}
  \caption{\textbf{Agent's behavior for the fourth metacognitive skill "ASSESS".}}
  \label{fig:snap}
\end{figure}

\textbf{The practice sessions :} During sessions 6 \& 7, children had to put into practice the skills they saw with the previous videos. More specifically, their task was to use a four-step learning loop (identify, guess, seek, assess) in order to acquire new and useful information about a given text. To do so, they used a web platform where they interacted with hand-crafted conversational agents representing the said metacognitive characters. The agents guide participants into using these skills by reminding them of their roles : once participants finish reading the text, the first controller appears (representing the IDENTIFY skill) and guides them into identifying a knowledge gap. After validating this step, the detective agent appears to guide them into formulating a hypothesis about it (GUESS skill), and so on. See Figure \ref{fig:snap} for an example of the interface and Appendix\ref{appendix:agents} for examples of the agents' utterances and snaps of the interactions with all four agents. The agents had predefined scripted behaviors and did not give any feedback regarding children's inputs.

During the first two cycles, and in order to help children get familiar with the task, the agents gave a list of propositions along with the description of their roles (i.e a list of uncertainties for the IDENTIFY agent, of guesses for the GUESS agent and of questions for the SEEK agent); children were free to either use the agents' propositions or to enter their own input. During the 6 remaining tasks, children did not have this help automatically and had to explicitly ask for it if they were blocked. 

In total, children had 8 texts, i.e. 8 learning cycles to complete following the four skills.

\subsection{Participants}
We recruited 15 CM1 students from a French primary school, aged between 8 and 10 years old, with 8 girls and 7 boys (see Appendix \ref{appendix:measures} for a detailed description of the participants' profiles).

The study was approved by the institute's ethics committee (certificate n°2019-23) and only started after having all participants parents' signed consents which contained the study goals, procedure and the data collected.

\section{Preliminary results}
\subsection{Accessibility of the workshop}
\subsubsection{Could children understand the content of the videos ?}
During each of the 4 introductory sessions, we wrapped up the session with a 6-item quiz about the content presented in the video to assess how much children were able to understand it. Results suggested indeed a good accessibility with a mean score of M=75.99\% and SD=23.62 for the four videos; details about the scores for each video can be found in Appendix \ref{appendix:scores-vid}. 

\subsubsection{Were children able to use the four metacognitive skills to perform curiosity-driven learning ?}
During the training with the web-based platform, children had to perform 8 curiosity-driven learning cycles using the four metacognitive skills. A complete cycle is one where children identify an uncertainty, generate a hypothesis for it, ask the relevant question about it and then decide whether or not they could find the answer they're looking for within a given list of propositions. Our results show that all children succeeded to complete at least four cycles during the training with a mean percentage of correct cycles completed of M=64.42\% and SD=20.31\%.

\subsection{Pre-post intervention effects}
\subsubsection{How did the workshop affect children's metacognitive sensitivity ?}
Children's metacognitive sensitivity reflects how accurate they are in judging their own level of knowledge, it is thus a facilitator for curiosity. See Appendix\ref{appendix:measures} for details about how we compute this measure.

\begin{figure}[ht]
  \centering
  \includegraphics[width=1\linewidth]{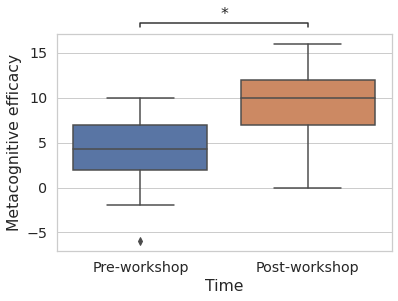}
  \caption{\textbf{Children's ability to judge the accuracy of their answers correctly was significantly enhanced after the workshop. Their understanding of the videos during the first part of the workshop had a significant effect on this progress (F(1,27)=20.67; p-value=0.0001).}}
  \label{fig:mc_eff}
\end{figure}

As shown in Figure \ref{fig:mc_eff}, results of the ANOVA test show indeed a significant enhancement in children's metacognitive efficiency (F(1,28)=8.2; p-value=0.007). Furthermore, and in investigating the role of our training on this enhancement, we run two ANCOVA tests with either score at the first (i.e. scores for videos) or the second (i.e. score of completion of learning cycles) part of the workshop as a co-variate. Results showed a significant effect for the understanding of the videos on children's metacognitive sensitivity (F(1,27)=20.67; p-value=0.0001) revealing that the higher declarative knowledge they gained from the videos, the more they had extensive gains in metacognitive sensitivity.


\begin{figure}[ht]
  \centering
  \includegraphics[width=1\linewidth]{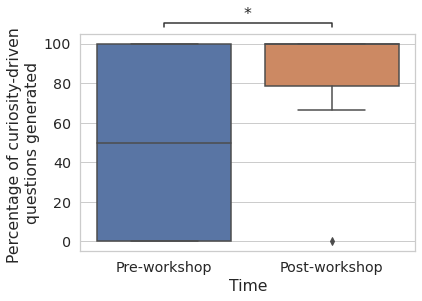}
  \caption{\textbf{Children's ability to ask curiosity-based questions during learning was significantly enhanced after the workshop. Their performance during the second part of the workshop had a significant effect on this progress (F(1,27)=4.21; p-value=0.05).}}
  \label{fig:qa}
\end{figure}

\subsubsection{How did the workshop affect children's curious question-asking skills ?}
Asking divergent questions is a primary expression of curiosity in children and can be seen as a behavioral measure for it~\cite{Aschner1963,alaimi2020}. To measure children's ability to formulate these questions, we use an offline test and expert annotations as described in Appendix\ref{appendix:measures}, before and after the workshop. 

Results showed a significant progress in this ability after the workshop, as shown in Figure \ref{fig:qa}. We also tested the effect of the training on this progress using ANCOVA tests. The tests reveal a significant effect of the performance during the practice sessions on the progress in time (F(1,27)=4.21; p-value=0.05) revealing that the higher they gained procedural knowledge from the practice sessions, the more extensive they had gains in question-asking performance.

\subsubsection{How did the workshop affect children's perception of curiosity ?}
Children's perception of curiosity is an important measure for us since it affects the individuals' curiosity-related behaviors. To assess this we use Post's standardized test~\cite{Post2015} (see details about the test in Appendix \ref{appendix:measures}).

Our results of the ANOVA test show no significant difference in this measure before and after the workshop (F(1,38)=0.37 with p-value=0.5). 

\section{Discussion}
The results of this study show the general accessibility of the two parts of the training and significant effects on behavioral curiosity-related measures. More specifically, we see that the first part which gives declarative knowledge about curiosity and metacognitive skills affects children's metacognitive sensitivity. While the second part that helps gain procedural knowledge these concepts affects their ability to ask curiosity-driven questions. Following Anderson and Krathwohl's taxonomy of learning~\cite{krathwohl2002}, these observations can be rather expected since metacognitive sensitivity require declarative knowledge about what one does and does not know while asking curious questions requires procedural knowledge (i.e. being aware of a missing information and then formulating a relevant question to reach it).  

However, we see no enhancement in participants' perception of curiosity with this first sample. This observation can be explained by the nature (i.e. digital interaction) and the short duration of the intervention (8 days). It is indeed well documented in social attitude literature that reliable and sustainable attitudinal changes often require longer interventions and/or with more realistic social interaction with teacher or between children~\cite{Cordova1996}.

\section{Conclusion, limitations and and future directions}
In this work, we contribute to the promotion of metacognition-based approaches to foster curiosity, while proposing novel activities. Despite these encouraging first results, more participants need to be recruited in order to replicate the findings and confirm the efficiency of the intervention. In a further step, we aim to leverage new technologies (e.g. AI-based systems) in order to have more powerful conversational agents and more fluid interactions with children. 

\section{Acknowledgments}
This work has been funded by the educational technologies start-up EvidenceB, the French National Association of Research and Technology (ANRT), and the ANR DeepCuriosity project. The authors thank Mr Benjamin Erben for helping create the content for the educational videos and Mrs Isabelle Coutand, the teacher who participated with her students in this study and the research team members who helped conduct the experiments in classes.

\begin{appendices}
\section{Curiosity framework with relation to the four metacognitive skills}
\label{appendix:cur-framework}
\begin{figure}[ht]
  \centering
  \includegraphics[width=1\linewidth]{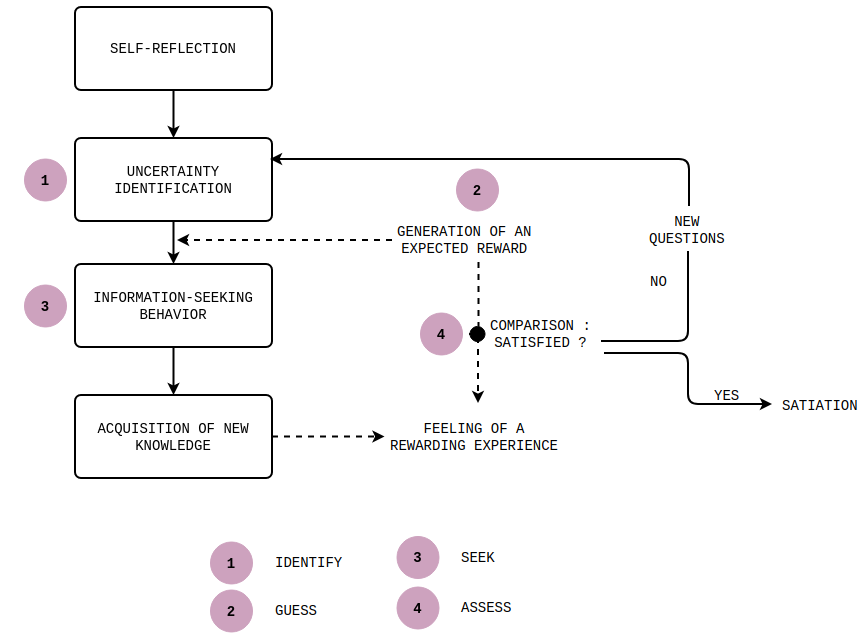}
  \caption{\textbf{Curiosity-driven learning framework and link with the metacognitive skills we propose to train during our intervention}}
  \label{fig:cur-framework}
\end{figure}

Taking inspiration from Murayama's curiosity-driven learning framework~\cite{murayama2019}, we link this aptitude with four essential metacognitive skills as seen in Figure ~\ref{fig:cur-framework}. 

\section{Pedagogical material for sessions 2--5 : videos content and metacognitive characters}
\label{appendix:mc-car}
\subsection{Pedagogical content for the animated videos}
The pedagogical content for the animated videos presenting declarative knowledge about curiosity and metacognitive-related skills are summarized below.

\begin{itemize}
    \item \textbf{Session 1 :} The goal is for children to understand what is motivation, its types and the importance of intrinsic motivation for learning. The videos introduces motivation as a fuel for learning : it gives the brain the goals it needs to stay engaged and active in the task. It introduces two types : extrinsic motivation that relies on external rewards from the environment such as money, etc and intrinsic motivation that is generated within the individual, following his interests, personal desires ... . We explain that the latter is more important for our brain to stay engaged in a task.

    \item \textbf{Session 2 :} The goal is for children to understand what is curiosity and its importance for learning. The video introduces curiosity as a form of intrinsic motivation. It presents its advantages following the three pillars for an efficient and enjoyable learning experience : 1) allows to modulate the individuals' memory, 2) guides their attention and 3) enhances their sense of self-competency.

    \item \textbf{Session 3 :} The goal is for children to understand what is metacognition and to introduce them to the four metacognitive skills affecting curiosity. The videos introduces metacognition as a set of skills that represents 'Thinking about thinking': it facilitates a longer-lasting learning and avoids the 'Illusion of Knowledge' trap. The video also introduces the four curiosity-related metacognitive skills that are important to learning : Observing and identifying uncertainty, predicting and generating guesses, pursuing knowledge gaps through question-asking and exploration and evaluating newly-acquired information. 

    \item \textbf{Session 4 :} The goal is for children to understand how to use metacognition to be more curious and be a better learner. The videos presents curiosity as a malleable skill that individuals can actually control, using their metacognitive skills. It emphasizes the need to use the observational skills to identify knowledge gaps that can poke curiosity. It also talks about how to control our own curiosity in order to learn tasks that are suited to our competency levels. And finally, the video encourages the use of the evaluation skills in order to be engage in further and longer-lasting learning.
    
\end{itemize}

\subsection{Description of the metacognitive characters}
The metacognitive characters reflect the four metacognitive skills involved in curiosity-driven learning; they were first introduced in the videos. They then accompany children in the practice sessions on the training platform. They were presented with the following roles :

\begin{itemize}
    \item \textbf{The controller} : when learning new content, it observes the task, reflects on its previous knowledge and chooses which uncertainty or missing information to pursue : reflects the IDENTIFY skill. 
    \item \textbf{The detective} : formulates educated guesses and makes predictions about the missing information to pursue : reflects the GUESS skill.
    \item \textbf{The explorer} : pursues the uncertainty by asking the relevant questions or by exploring the relevant resources : reflects the SEEK skill.
    \item \textbf{The second controller} : evaluates whether the inquiry resulted in learning progress with respect to the initial uncertainty : reflects the ASSESS skill. 
\end{itemize}

\section{Material for sessions 6--7 : The training platform}
\label{appendix:agents}
\begin{figure}[ht]
  \centering
  \includegraphics[width=1\linewidth]{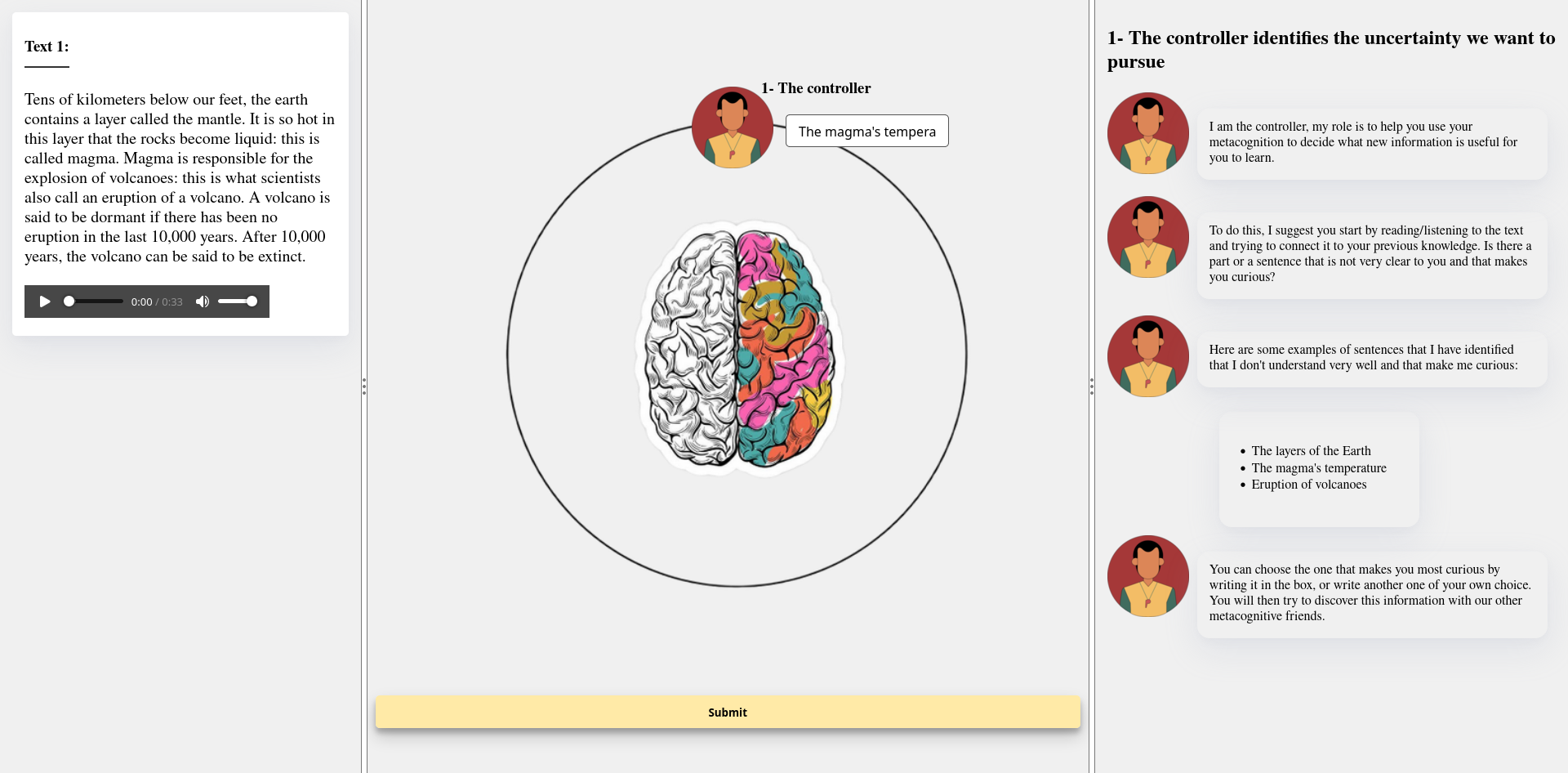}
  \caption{\textbf{Agent's behavior for the first metacognitive skill "IDENTIFY".}}
  \label{fig:snap1}
\end{figure}

\begin{figure}[ht]
  \centering
  \includegraphics[width=1\linewidth]{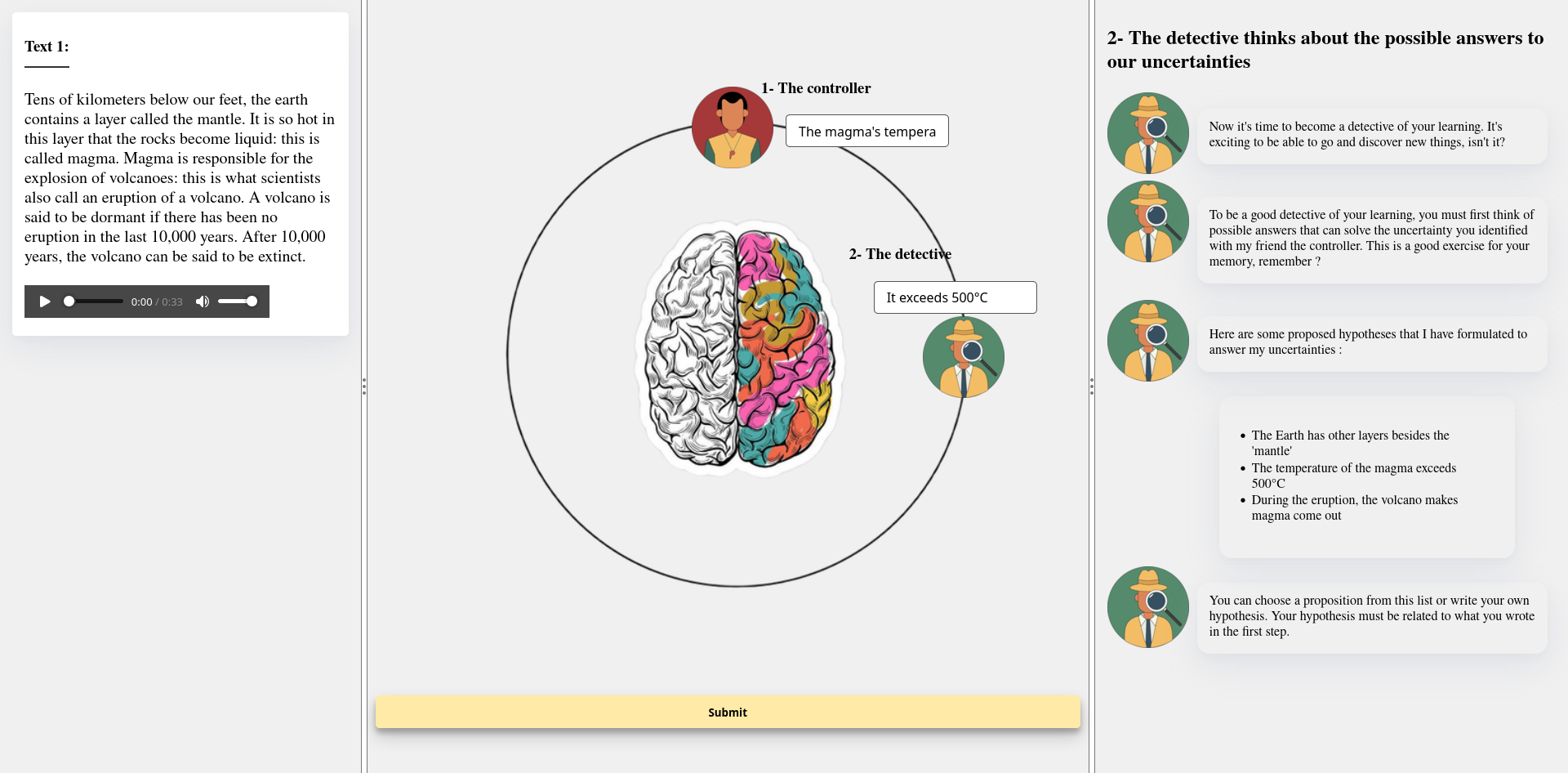}
  \caption{\textbf{Agent's behavior for the second metacognitive skill "GUESS".}}
  \label{fig:snap2}
\end{figure}

\begin{figure}[ht]
  \centering
  \includegraphics[width=1\linewidth]{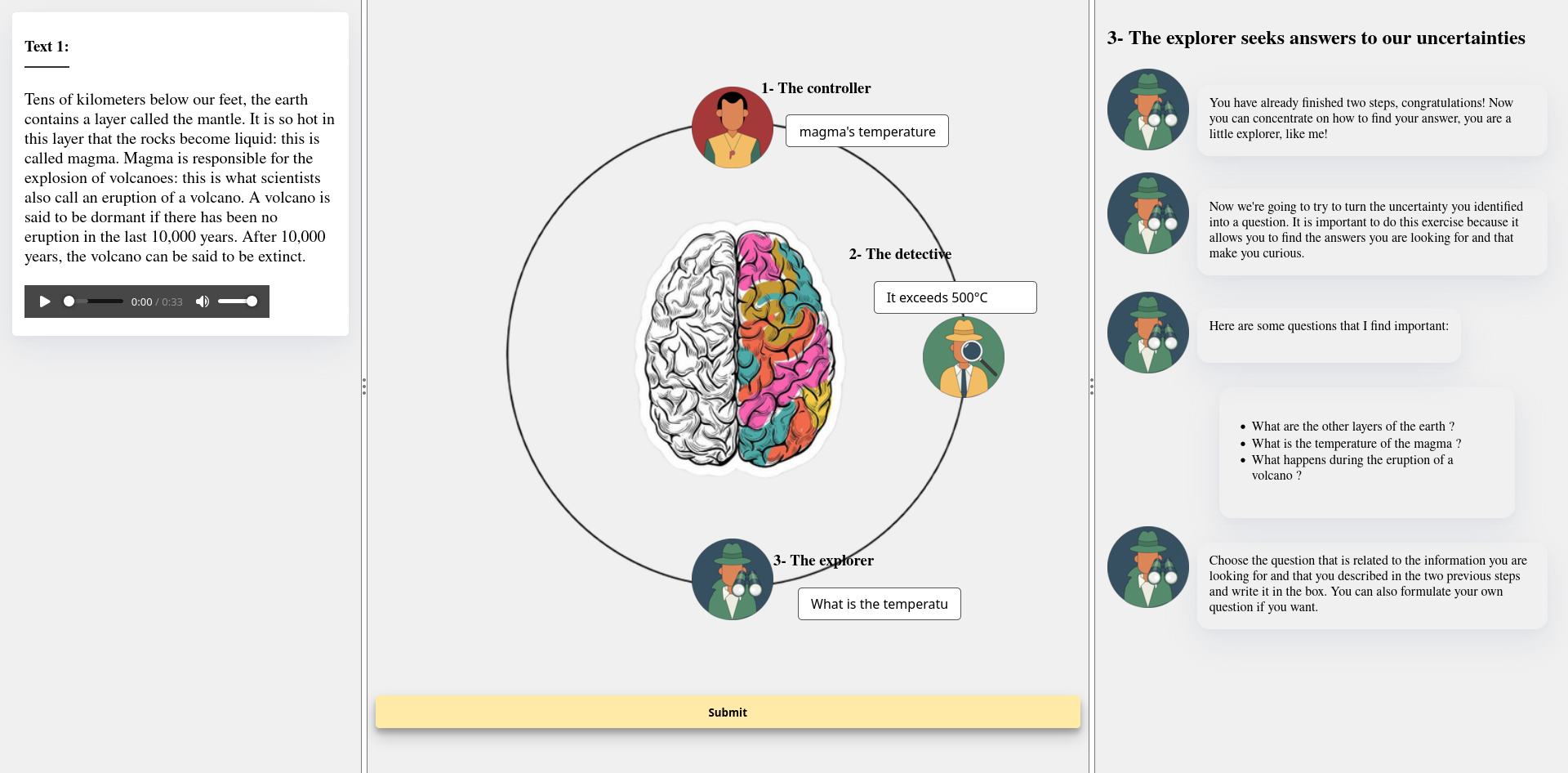}
  \caption{\textbf{Agent's behavior for the third metacognitive skill "SEEK".}}
  \label{fig:snap3}
\end{figure}

Examples of the utterances of the agents representing the four curiosity-related metacognitive skills can be seen in the platform snaps (Figures \ref{fig:snap1} -- \ref{fig:snap3}).


\section{Details for data collection procedure and measures}
\label{appendix:measures}
\subsection{Profile measures}
We were left with 15 participants, the description of their profiles can be found in Table \ref{Tab:mesures}.

\begin{table}
    \caption{\textbf{Profile measures for the participants}}
  \label{Tab:mesures}
  \begin{tabular}{ccl}
    \toprule
    Measure&Mean $\pm $ std\\
    \midrule
    Age & 9.05 $\pm $ 0.53\\
    Device use frequency & 30.88 $\pm $ 6.66\\
    Curiosity trait & 27.6 $\pm $ 4.9\\
    Perception of curiosity & 34.05 $\pm $ 8.8\\
  \bottomrule
\end{tabular}
\end{table}

For these profile measures, we used the following standardized questionnaires :

\begin{itemize}
    \item \textbf{Curiosity trait :} 
To assess children's curiosity trait, their parents answered Litman's questionnaire of curiosity~\cite{Litman2004}.
    \item \textbf{Perception of curiosity :}
This is an important measure for us as we think that one of the relevant brakes that can keep children from asking questions in the classroom is their negative perception of this behavior. To assess children's perception of this, we use Post's validated CIAC questionnaire~\cite{Post2015}. This questionnaire contains items to assess children's general attitude towards curiosity and asking questions, their perception of its importance and their negative ideas about it, if any.
\end{itemize}

\subsection{Behavioral curiosity-related measures}
\textbf{Metacognitive sensitivity :}
This is an important measure for us as recognizing uncertainties is a key step to engage in curious learning. We measure this using a general knowledge quiz where participants report their confidence levels in their answers. We then compute the difference between the 'hit rates' i.e. the number of times their confidence judgment corresponds to their response accuracy) and the 'false alarms' (i.e. number of times they incorrectly judge their answer)~\cite{fleming2017}.

\textbf{Ability to ask curious questions :} 
Asking divergent questions is a primary expression of curiosity in children and can be seen as a behavioral measure for it~\cite{Aschner1963,alaimi2020}. Divergent questions bring new information and require more complex reasoning and process such as linking two ideas, making hypotheses, etc. To assess this aspect, we use an offline test inspired from creativity studies such as in~\cite{diakidoy2002}. This test consists of giving participants a short text to read then asking them to write as many questions about it as they can, within 2 minutes. We then count the total number of the correct questions generated during the test and compute the percentage of the divergent questions amongst them using human experts annotations. A question is accepted if it its indeed, a question not a statement, and is semantically related to the text. If a question is repeated, it is only taken into account once. 

A question is counted as divergent if its answer is explicitly stated in the text. This is based on Gallagher's classification ~\cite{Aschner1963}. \textbf{Example:} For the text in the screenshots above, the question “What is the temperature of the magma?” is considered to be divergent, whereas "Where is the mantle situated?" is not as the answer to it is explicitly stated in the text.

This process of data coding was performed by one of the researchers that led this study. All data was anonymized: the coder could only see the identifiers that children were given randomly at the beginning of the intervention.

\subsection{Learning experience measures}
\textbf{Motivation :}
The participants’ motivation to use the proposed platform was assessed via two questionnaires: the general motivation scale \cite{Cordova1996} and Vallerand's types of motivation scale \cite{vallerand1989}. 

The general motivation scale was used to investigate the potential short-term motivation. It contains one sub-scale for evaluating participants' motivation to reuse the platform in the future, one sub-scale for evaluating their perceived competence and one sub-scale for assessing the degree of preference with respect to a favourite school activity.

On the other hand, Vallerand's scale was used to probe intrinsic and extrinsic motivational mechanisms in the educational settings we had. It is composed of three sub-scales that differentiate: intrinsic motivation, extrinsic motivation and amotivation.

\textbf{Task load :}
The participants' behavior was also evaluated in terms of the subjective workload they experienced during the intervention. For this, we used the NASA-TLX workload multi-dimensional scale developed in \cite{Hart1988}. Information about the intensity of six workload-related factors are used in order to estimate a reliable measure of workload: mental demand, physical demand, temporal demand, performance, effort and frustration. 

\section{Additional results}
\subsection{Details about the scores of the four videos}
\label{appendix:scores-vid}
The details of the accessibility of the four videos are like the following : video 1 concerning the definition of motivation and its types (M=71.79\% and SD=21.48) ; video 2 concerning the definition of curiosity and its importance to learning (M=84.05\% and SD=21.6) ; video 3 concerning the definition of metacognition and the four skills involved in curiosity (M=74\% and SD=25.49) ; and video 4 concerning how metacognition helps curiosity and learning (M=74.13\% and SD=23.62).

\subsection{How motivated were children to do the training ?}
Children's general motivation to use the platform was rather high (M=37.72 and SD=10.24) for a maximum score of 54. For the types of motivation, children were more intrinsically than extrinsically motivated to do the training (M=76.78\% and SD=30.19 for intrinsic motivation and M=66.66\% and SD=28.7 for extrinsic motivation). See Appendix~\ref{appendix:measures} for details about the scales used to measures these data.

\subsection{How demanding was the training for children ?}
Results of the Nasa-tlx scale~\cite{Hart1988} revealed that participants did not perceive the tasks as very demanding : M=27.57 and SD=11.17 for a total possible score of 60. See details about this scale in Appendix~\ref{appendix:measures}. 

\end{appendices}
\bibliographystyle{abbrv}
\bibliography{bibfile}
\end{document}